\documentclass[article,aip,pof,groupedaddress,showpacs,preprint,english]{revtex4-1}
\usepackage[T1]{fontenc}
\usepackage[utf8]{luainputenc}
\usepackage{textcomp}
\usepackage{amsmath}
\usepackage{amssymb}
\usepackage{graphicx}
\usepackage{color}
\makeatletter

\newcommand{\lyxmathsym}[1]{\ifmmode\begingroup\def\b@ld{bold}
  \text{\ifx\math@version\b@ld\bfseries\fi#1}\endgroup\else#1\fi}

\providecommand{\tabularnewline}{\\}

\@ifundefined{textcolor}{}
{%
 \definecolor{BLACK}{gray}{0}
 \definecolor{WHITE}{gray}{1}
 \definecolor{RED}{rgb}{1,0,0}
 \definecolor{GREEN}{rgb}{0,1,0}
 \definecolor{BLUE}{rgb}{0,0,1}
 \definecolor{CYAN}{cmyk}{1,0,0,0}
 \definecolor{MAGENTA}{cmyk}{0,1,0,0}
 \definecolor{YELLOW}{cmyk}{0,0,1,0}
 }

\usepackage{latexsym}%\newcommand{\W}{13cm}

\newcommand{\W}{11cm}
\usepackage{babel}
\begin{document}
% Use the \preprint command to place your local institutional report
% number in the upper righthand corner of the title page in preprint mode.
% Multiple \preprint commands are allowed.
% Use the 'preprintnumbers' class option to override journal defaults
% to display numbers if necessary
%\preprint{}

%Title of paper

\title{Oscillations  and translation of a free cylinder  in a
confined flow}
%%%%%%%%%%%%
% repeat the \author .. \affiliation  etc. as needed
% \email, \thanks, \homepage, \altaffiliation all apply to the current
% author. Explanatory text should go in the []'s, actual e-mail
% address or url should go in the {}'s for \email and \homepage.
% Please use the appropriate macro foreach each type of information
%%%%%%%%%%
% \affiliation command applies to all authors since the last
% \affiliation command. The \affiliation command should follow the
% other information
% \affiliation can be followed by \email, \homepage, \thanks as well.
\author{Maria Veronica D'Angelo}
\email{vdangelo@fi.uba.ar}
\affiliation{Grupo de Medios
Porosos, Facultad de Ingenier\'{\i}a, Paseo Colon 850, 1063, Buenos
Aires (Argentina)}
\affiliation{Univ Pierre et Marie Curie-Paris 6, Univ Paris-Sud, CNRS, F-91405.
  Lab FAST, B\^at 502, Campus Univ, Orsay, F-91405 (France).}
\author{Jean-Pierre Hulin}
\email{hulin@fast.u-psud.fr}
\affiliation{Univ Pierre et Marie Curie-Paris 6, Univ Paris-Sud, CNRS, F-91405.
  Lab FAST, B\^at 502, Campus Univ, Orsay, F-91405 (France).}
%%%%%%%%%%%%%%%%%%%%%
\author{Harold Auradou}
\email{auradou@fast.u-psud.fr}
\affiliation{Univ Pierre et Marie Curie-Paris 6, Univ Paris-Sud, CNRS, F-91405.
  Lab FAST, B\^at 502, Campus Univ, Orsay, F-91405 (France).}
%\email[]{Your e-mail address}
%\homepage[]{Your web page}
%\thanks{}
%\altaffiliation{}

%Collaboration name if desired (requires use of superscriptaddress
%option in \documentclass). \noaffiliation is required (may also be
%used with the \author command).
%\collaboration can be followed by \email, \homepage, \thanks as well.
%\collaboration{}
%\noaffiliation

\date{\today}
\begin{abstract}
An oscillatory instability   has been observed
 experimentally on  an horizontal cylinder free to move and rotate 
between two parallel vertical walls of distance $H$; its characteristics
differ both from  vortex shedding driven oscillations
and from those of tethered cylinders in the same
geometry. 
The vertical motion of the cylinder, its rotation about its axis and
its transverse motion across the gap  have been investigated
 as a function of  its diameter $D$, its density $\rho_{s}$,
of the mean vertical velocity $U$ of the fluid and of its viscosity.
For a blockage ratio $D/H$ above $0.5$ and a Reynolds number $Re$ larger
then $14$, oscillations of the rolling angle of the cylinder about its axis and of
its transverse coordinate in the gap are observed together with periodic variations
of the vertical velocity. Their frequency $f$ is  the same for the sedimentation of the cylinder in
a static fluid ($U = 0$) and for a non-zero mean flow ($U \neq 0$).
 The Strouhal number $St$ associated to the oscillation varies as $1/Re$ with : 
 $St.Re = 3\pm0.15$.  The corresponding period $1/f$ is then independent of $U$ and 
corresponds to a characteristic viscous diffusion time over  a distance  $\sim D$, implying 
a strong influence of the viscosity. These characteristics  differ  from those of
vortex shedding and tethered cylinders for which $St$  is instead roughly constant  
with $Re$ and higher than  here. 
%%%%%%%%%%%%%%%%%%
\end{abstract}
% insert suggested PACS numbers in braces on next line

% insert suggested keywords - APS authors don't need to do this
%\keywords{Rheology, Flow instability, porous media}

%\maketitle must follow title, authors, abstract, \pacs, and \keywords

% body of paper here - Use proper section commands
%%%%%%%%%%%%%%%%%%%%%%%%%%%%%%%%%%

\maketitle

\section{Introduction} \label{intro} 
%%%%%%%%%%%%%%%%%%%%%%%%%%
The influence of confinement on the motion of a cylinder facing  a 
flow is relevant to many applications like  the transport of particles or fibers in slits or the 
development and localization of  bio films inside pores \cite{Rusconi2010,Autrusson2011}.
Many studies have been devoted to this problem but dealt mostly with
the determination of the forces on the cylinders (for instance, when they were left
free to rotate or eccentered in a stationary flow).

In the studies of the hydrodynamical transport of confined 
cylinders~\cite{Dvinsky87a,Dvinsky87b,Eklund1994,Hu95},
it has  usually been assumed that, in the absence of vortex shedding, the
motion of the cylinder is steady: the cylinder translates with constant
velocity and, in some cases,  rotates with a constant angular velocity and at
a fixed transverse distance  from the mid plane of the gap.

Following these views, non-stationary flows would  only appear  at Reynolds numbers, $Re$, 
above  the vortex shedding threshold. The present work demonstrates
 instead, at lower Reynolds numbers, 
a periodic non-stationary transport regime due to another type of flow instability strongly
influenced  by the viscosity.

Early studies of the torque and  drag forces on a cylinder facing a flow
have been performed  in the Stokes regime or at relatively small Reynolds
numbers.  For particles placed in the centre of the channel, Faxen \cite{Faxen1946}
derived the expression of the drag for a confinement $D/H$ less then
$0.5$;  the case of higher confinements has been recently considered by Ben
Richou and co workers \cite{Richou2004,Richou2005}. An eccentered  cylinder 
experiences in addition a positive torque decreasing sharply in
the vicinity of the walls~\cite{Dvinsky87a,Dvinsky87b,Hu95,Eklund1994};
 for  a cylinder translating closely along a wall or held fixed in a Poiseuille flow, 
 this torque tends to generate a rotation 
of sign opposite to that of contact rolling. This results from the backflow
near the second wall and has a sizable influence on the force
distribution on the cylinder \cite{Champmartin2007}.

The displacement of a free cylinder released from an eccentric
position inside a vertical gap has been computed by Hu~\cite{Hu95}
 for three values of the  Reynolds number $Re$. 
For $Re \le 5$, the cylinder reaches a  final stable transverse
position  in the middle of the gap. For $Re \simeq 100$,  instead, an off-axis cylinder 
rotates in the direction opposite to the previous one, resulting in a lift force oriented
away from the axis: this was accounted for by  the appearance of a recirculation 
zone~\cite{Juarez2000}. However, as the cylinder approaches one of the walls, the 
recirculation zone recedes because of the interaction between the wake and the 
wall boundary layer. The rotation and the lift force then change sign again, so that a stable
off-axis position  is finally found. 

Such observations are made at a Reynolds number close to
 the periodic vortex shedding regime \cite{Sahin2004}
and are consistent with the conclusions of Zovatto and Pedrizzetti
\cite{Zovatto2001}  for a non rotating cylinder. Above the
critical Reynolds number, vortex shedding may induce vibrations
of frequency and amplitude depending on the mechanical properties
of the system \cite{Delangre2010,Shiels2001,Williamson2008}.

More recently, Semin et al \cite{Semin2011} observed that a  tethered
cylinder placed in a Poiseuille flow between vertical parallel planes oscillates  spontaneously  at 
Reynolds numbers below the threshold for vortex shedding: unlike in the present
case, both the vertical and rolling motions of the cylinder were blocked.

The present work deals with an  horizontal cylinder free to translate and rotate inside the gap of a
vertical  Hele Shaw cell. Either this cylinder sediments in a stationary fluid or is submitted to a
vertical Poiseuille flow ({\it i.e.} transverse to its axis): the relative velocity of the 
cylinder and of the fluid is always below the threshold for vortex shedding.
The transverse and vertical components of 
the motion of the cylinder and its rotation about its axis are studied: 
 the influence of physical parameters such as the diameter and 
density of the cylinder and the viscosity of the fluid and of hydrodynamical variables
like the flow velocity is particularly investigated.
%%%%%%%%%%%%%%%%%%
\section{Description of the experimental setup}
%%%%%%%%%%%%%%%%%%%%%
%%%%%%%%%%%%%%%%%%%%%%%%%%%%%%%%%%%%%%%%
\begin{figure}
\includegraphics[width=\W]{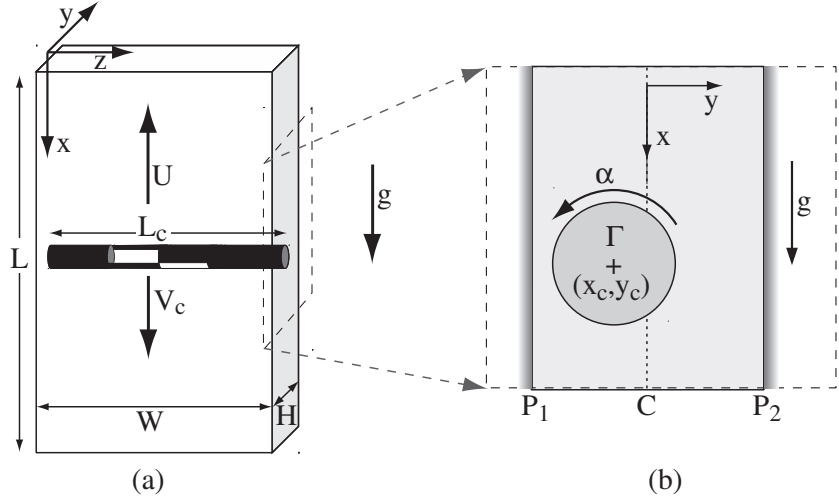} 
\caption{Experimental setup. a) Front view - $U$: mean flow velocity, $V_{cx}$:  vertical
component of the cylinder velocity. b) Side view - $y_{c}$:
transverse position of the center of mass of the cylinder, $\Gamma$ :
Torque.}
\label{fig:exp} 
\end{figure}
%%%%%%%%%%%%%%%%%%%%%%%%%%%%
The experimental setup consists of a Hele Shaw cell placed vertically.
Its height, width and aperture are respectively $L = 350$, $W=100$ and
$H=3\ \mathrm{mm}$. The vertical sections of the cell  have a Y-shape in their upper part; the upper end of 
the cell is at the bottom of a rectangular bath with a slit allowing for the flow of the fluid
and the insertion of the cylinders.  An upward flow  may be  imposed by a  gear pump: the  fluid velocity 
 $U$ is counted in this case as negative.

%%%%%%%%%%%%%%%%%%%%%%%%%%%%
\begin{table}
\begin{tabular}{lll}
Fluids  & $\rho_{f}$ ($\mathrm{g/cm}^{3}$)  & $\mu$ (mPa.s)\tabularnewline
\hline 
$WG$  & $1.05$  & $1.56$\tabularnewline
$N1$  & $0.998$  & $1.11$\tabularnewline
$N2$  & $0.998$  & $2.20$\tabularnewline
\end{tabular}
\caption{Physical properties of the fluids. density: $\rho_{f}$, viscosity:
$\mu$. Temperature $23^o C$. $N1$ and $N2$ correspond
to natrosol solutions at respectively $1$ and $2\ \mathrm{g.l}^{-1}$. $WG$ refers
to a glycerol solution containing $20\%$ in weight of glycerol.}\label{tab:tab0} 
\end{table}
%%%%%%%%%%%%%%%%%%%%%%%%%%%%%%%%%
Table~\ref{tab:tab0} lists the characteristics of the fluids used in the experiments; 
the viscosity is measured using a Contraves Low Shear-30 rheometer.
 In this study, the natrosol concentration
is sufficiently low so that the fluids can be considered as Newtonian: at a given temperature, their viscosity
is determined by the natrosol concentration (and increases with it).
For shear rates ranging from $0.2$ to $118\ \mathrm{s}^{-1}$, the viscosity
(see Tab.~\ref{tab:tab0})  of the two natrosol solutions is indeed found to be
constant (within $\pm 0.04\ \mathrm{mPa.s}$). The
density and temperature of the solutions are measured prior to any set
of experiments.

 The cylinders are made of PMMA (density $\rho_{s}=1.20\ \mathrm{g.cm}^{-3}$) or
of carbon ($\rho_{s}=1.54\ \mathrm{g.cm}^{-3}$); their diameter $D$ ranges from $1.1$ 
to $2.1\, \mathrm{mm}$. Their length $L_c$ is smaller than but as close as possible 
to the internal  width $W$ of the cell.  Initially, they are placed
in the upper bath with their principal axis horizontal and one lets them move down
into the Y-shaped zone by reducing the flow rate $Q$. 
Then, $Q$ may  be adjusted so that   the cylinder remains at a fixed level either at rest (state
$0$) or oscillates about its principal axis (state $1$). Then,  one may reduce $Q$ (sometimes to zero)  in order
to   analyze   the motion of falling  cylinders;  in a part of the experiments,
$Q$ is increased again after the cylinder has reached the bottom of the 
cell for studying  its upward motion ($V_{cx} < 0$).

The displacement of the cylinder is monitored by two  cameras triggered synchronously;
they image respectively the displacements in the
plane $(x,z)$ of the Hele Shaw cell  and in the plane $(x,y)$ of the gap (the axis $y = 0$
is in the midplane between the walls). 
Processing digitally the two sets of images gives first the instantaneous coordinates $(x_c, y_c)$ of the
 center of mass of the cylinder in the  $(x,z)$ and  $(x,y)$ planes. The  angle $\theta$ 
 between its axis and the horizontal is also determined from the instantaneous location of its two
 ends in the  $(x,z)$ plane. 
 In order to analyze the 
rotations of the cylinder around its axis,
its length  is divided into $4$ domains of equal size. 
The two outside
parts are painted in black and two black staggered stripes parallel to the axis are painted
on the  central portions. 
The rotation
 about the axis is  analyzed from the variation with time of 
 the local vertical distance between each of  the stripes and the 
 principal axis of the rod: this allows one to determine the rotation angle
  $\alpha$ and, therefore, the corresponding angular velocity $\dot{\alpha}$.
  
 The Reynolds number $Re$ is defined by:
 %%%%%%%%%%%%%%%%%
 \begin{equation}
  Re=\rho_{f} |U^{*}| (H-D)/(2\mu),
  \label{Reynolds}
  \end{equation}
 %%%%%%%%%%%%%%%%% 
in which $(H-D)/2$ is the width of the gap between the walls and
the cylinder (when $y_c = 0$), $U^{*}$ is the mean instantaneous vertical fluid velocity in this gap, $\rho_{f}$
and $\mu$ are the fluid density and viscosity. The flow in the gap
combines the component imposed by the pump and that induced
by the vertical displacement of the cylinder. The velocity $U^{*}$ is defined as : 
%%%%%%%%%%%
\begin{equation}
U^{*} = \frac{H}{H-D} U - \frac{D}{H-D} V_{cx},
\end{equation}
%%%%%%%%%%%%%%%%%%%%%
in which $U$ is the mean velocity  far  from the 
cylinder and $V_{cx}$ is the vertical component of its velocity.

%%%%%%%%%%%%%%%%%%%%%%%%%%%%%%%%%%%%%%%%
\begin{figure}
\includegraphics[width=\W]{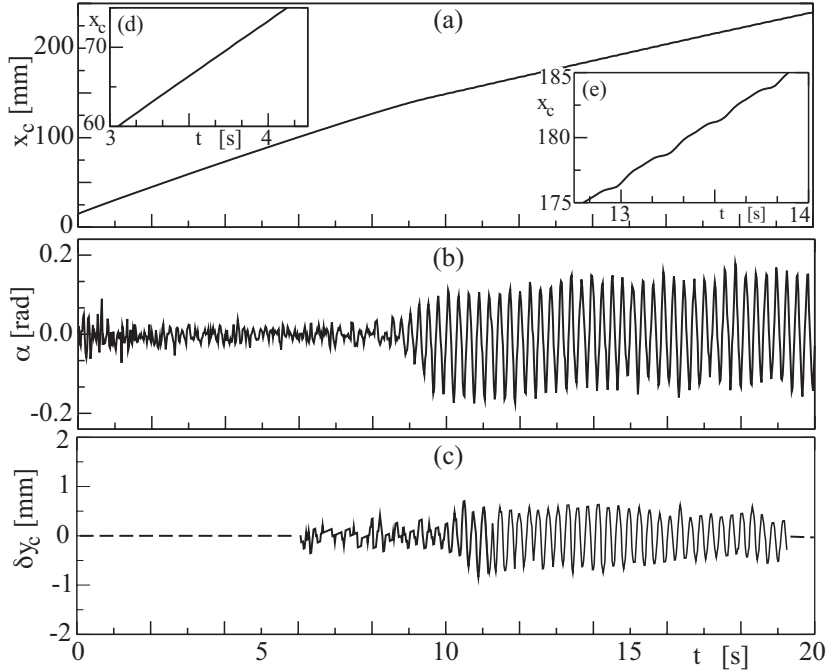} \caption{Experimental measurements obtained 
using solution $N1$ (see Tab.~\ref{tab:tab0}) with $D=1.45\ \mathrm{mm}$ ($D/H=0.48$) 
and $U = -10.55\ \mathrm{mm.s}^{-1}$. $(a)$ $x_{c}$
vs time $t$ ($s$). $(b)$ $\alpha$ vs $t$. $(c)$ $\delta y_c$
vs $t$.}
\label{fig:fig0} 
\end{figure}
%%%%%%%%%%%%%%%%%%%%%%%%%%%%%%%%%%%
Figure~\ref{fig:fig0} displays  experimental results obtained  using
  a PMMA  cylinder of diameter $D=1.45 \, \mathrm{mm}$ and   solution $N1$. 
 Two distinct regimes are observed:

- at the beginning of this experiment ($t \lesssim  8\ \mathrm{s}$), the cylinder
is located midway between the two vertical walls ($y_c = 0$ in Fig.\ref{fig:fig0}c);
it does not roll about its principal axis ($\alpha=0$ in Fig.\ref{fig:fig0}b)
and falls at a constant velocity (a linear regression gives $V_{cx} = 13.8\ \mathrm{mm.s}^{-1}$).

- After about $8\ \mathrm{s}$ the motion of the cylinder suddenly shifts to an oscillating regime:  both 
 the angle $\alpha$ of the cylinder about its axis and 
the deviation $\delta y_c$ from the mean transverse position in the gap oscillates with a 
well defined frequency. At the same time, the vertical translation velocity  $V_{cx}$ 
 drops by more than $35 \%$. 
These two  regimes are  discussed in detail  in sections~\ref{sec:RegimeT}
and \ref{sec:RegimeS} below.
%%%%%%%%%%%%%%%%%%%%%%%%%%%%%
\section{Mean vertical translation velocity for low Reynolds numbers
and moderate confinement.}
\label{sec:RegimeT} 
%%%%%%%%%%%%%%%%%%%%%%
In the present section we are interested in the   value of the vertical 
velocity $V_{cx}$ of the cylinder, averaged  over a time larger  than
 the period of the oscillations (if present) but short enough to avoid
the influence of global variations. In this section, ``velocity''  
always refers to such an average:  for instance in the case of Fig.~\ref{fig:fig0},
separate averages are computed  before and after the appearance of the 
oscillations.
The vertical velocity is  obtained from  the equilibrium condition
of the gravity and the vertical drag force $F_x$ (averaged over the same time lapse): 
%%%%%%%%%%%%%%%%%
\begin{equation}
m g + {F_x} = 0;
\end{equation}
%%%%%%%%%%%%%
here, $m = \pi \, \Delta\rho (D/2)^{2}$ is the reduced mass per unit 
length (with $\Delta\rho = \rho_{s} - \rho_{f}$). For a cylinder moving
 at a constant velocity $V_{cx}$ in a fluid flowing at
a constant mean velocity $U$ away from the cylinder, the drag
may be written in the low Reynolds number limit and when the cylinder does not rotate : 
%%%%%%%%%%%%%%%%%%%%
\begin{equation}
{F_x} = - \lambda_{s} \mu {V}_{cx} + \lambda_{p} \mu {U};\label{eq:drag}
\end{equation}
%%%%%%%%%%%%%%%%%%%%
the parameters $\lambda_{s}$ and $\lambda_{p}$ reflect
the influence of the geometrical confinement. For a long cylinder ($L_c \simeq W \gg D$
and $L_c \gg H$), $\lambda_{s}$ and $\lambda_{p}$ are only functions of the ratio of the cylinder
diameter $D$ and of the cell aperture $H$~\cite{Semin2009}. The vertical velocity of the cylinder is then: 
%%%%%%%%%%%%%%%%%%%%%%%%%%
\begin{equation}
{V}_{cx} = \displaystyle{\frac{\lambda_{p}} {\lambda_{s}}} {U} + {V}_{cx}^{0}, \label{eq:theory}
\end{equation}
%%%%%%%%%%%%%%%%%%%%%%%%%%%%
 in which 
%%%%%%%%%%%%%%%%%
\begin{equation}\label{vsed}
 {V_{cx}^{0}} =  m {g} / (\lambda_{s} \mu)
 \end{equation}
%%%%%%%%%%%%%%%%%%
 is the velocity of the cylinder with no applied flow ($U = 0$).

%%%%%%%%%%%%%%%%%%%%%%%%%%%
\begin{figure}[htbp]
\includegraphics[width=\W]{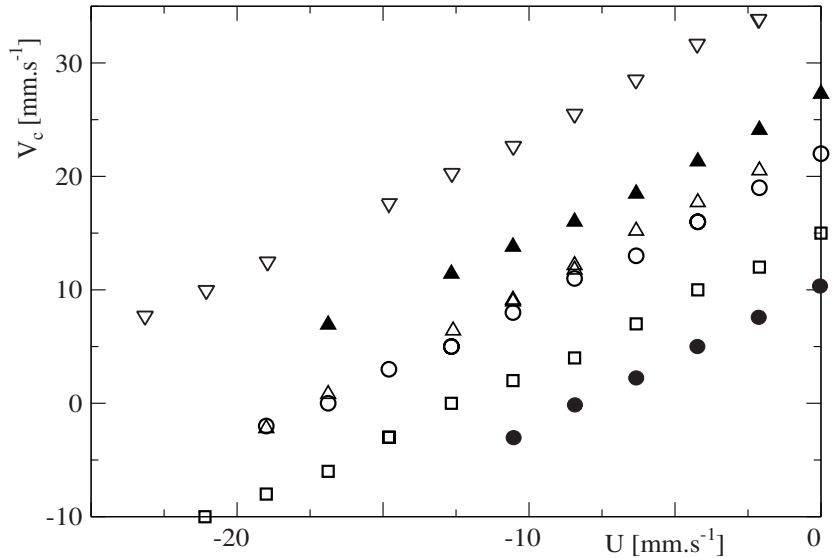} \caption{Vertical velocity $V_{cx}$ of   cylinders 
of diameter $D$ in solutions $N1$ or $N2$ (see Tab.~\ref{tab:tab0}) 
as a function
of the fluid velocity $U$. Open symbols: oscillating cylinders; solid symbols: no oscillations. 
PMMA cylinders - ($\triangle$), ($\blacktriangle$):  $D = 1.45\ \mathrm{mm}$ ($D/H=0.48$), $N1$; 
 ($\circ$), ($\bullet$):  $D=1.63\ \mathrm{mm}$ ($D/H=0.54$), $N1$ for ($\circ$) and $N2$ for ($\bullet$);  ($\square$):  $D=2.1\ \mathrm{mm}$  ($D/H=0.7$), $N1$. Carbon cylinder - ($\triangledown$):  $D = 1.45\ \mathrm{mm}$ ($D/H=0.48$),
$N2$.}
\label{fig:RegimeT-fig1} 
\end{figure}
%%%%%%%%%%%%%%%%%%%%%%%%%%%
Fig.~\ref{fig:RegimeT-fig1} displays the variation of the velocity
$V_{cx}$ of the cylinder with the mean flow velocity $U$ in the oscillation
and stationary regimes. As predicted
by Eq.~(\ref{eq:theory}) for viscous flows, $V_{cx}$
increases linearly with $U$ in both cases. 
For a same cylinder diameter ($D = 1.45\ \mathrm{mm}$) 
and a same fluid ($N1$) the velocity $V_{cx}$  in the oscillation regime is  lower than 
in the stationary one at all velocities $U$ as suggested above
(($\triangle$) and ($\blacktriangle$) symbols in Fig.~\ref{fig:RegimeT-fig1}); the slope of the 
variation with $U$ is also  slightly larger in the oscillation regime.

As also observed by Dvinsky and Popel \cite{Dvinsky87a},
 the sedimentation velocity $V_{cx}^{0}$ in a stationary fluid ($U = 0$) decreases with
the confinement: more generally, at a same velocity $U$, the cylinder velocity $V_{cx}$  is always slightly lower
 for $D/H = 0.54$ than for $D/H = 0.48$ and significantly lower for $D/H = 0.7$ 
(($\triangle$), ($\circ$) and ($\square$) symbols in Fig.~\ref{fig:RegimeT-fig1}).
This would have been the opposite for cylinders falling in a tank
of size much larger than their diameter  ($D/H\ll1$): in this unconfined case, the sedimentation 
rate increases with $D$ because the mass $m$ per unit of length varies faster (as $D^{2}$)
than the drag force.

In the confined case, instead,
Ben Richou et al.~\cite{Richou2005} found numerically in the lubrication approximation that the geometrical
factor $\lambda_{s}$ increases  like $D^{5/2}$ for  $D/H > 0.1$. 
Combining the  variations of $m$ and  $\lambda_{s}$ in Eq.~\ref{vsed}, the velocity 
$V_{cx}^{0}$  must then decrease with $D$ (or equivalently with $D/H$) as is  indeed observed.

Finally, the experiments confirm that increasing the fluid viscosity for a given cylinder 
reduces the value of $V_{cx}$ and result in a transition from an oscillation to a stationary regime
(($\circ$) and ($\bullet$) symbols in Fig.~\ref{fig:RegimeT-fig1}). For a significantly larger cylinder density,
$V_{cx}$ increases strongly, even for more viscous fluids (($\triangledown$)  
symbols in Fig.~\ref{fig:RegimeT-fig1}).

%%%%%%%%%%%%%%%%%%%%%%%%%%%%%%%%%%%%
\begin{figure}[htbp]
\includegraphics[width=\W]{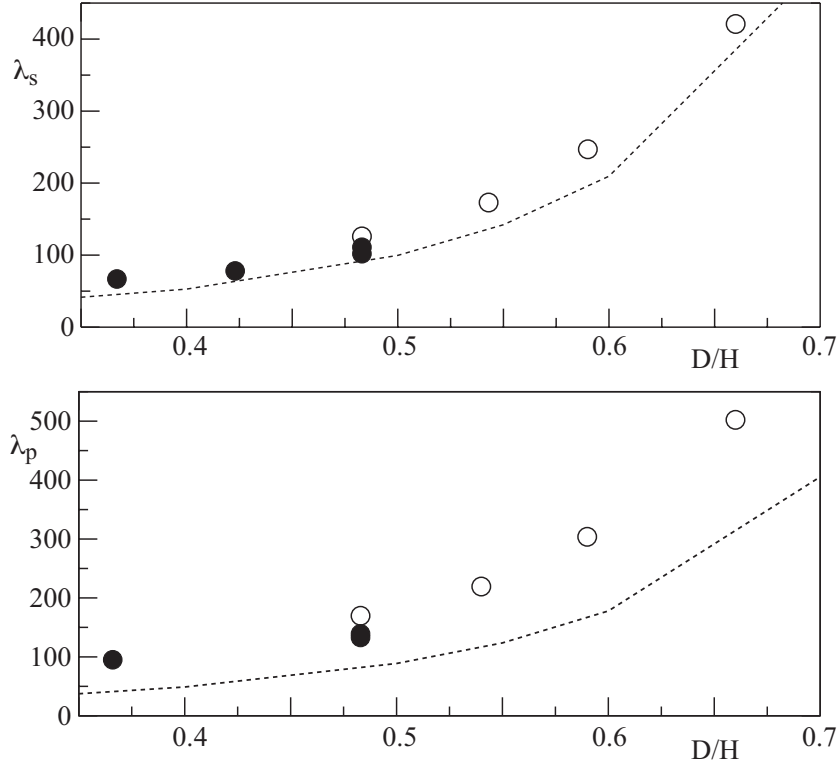} \caption{Top: Variations of $\lambda_{s}$ (top) and $\lambda_{p}$ (bottom) vs $D/H$. ($\circ$),($\bullet$) : experimental values of $\lambda_{s}$ (top) and $\lambda_{p}$
 respectively with and without oscillations; dashed lines: numerical data from ref.~\cite{Richou2005} (for $\lambda_{s}$) and ref.~\cite{Richou2004} (for $\lambda_{p}$).}
\label{fig:RegimeS-fig2} 
\end{figure}
%%%%%%%%%%%%%%%%%%%%%%%%%%%%%%%%%
In these experiments, the factor $\lambda_s$ is deduced by means of Eq.~\ref{vsed} from 
the experimental data for  $V_{cx}^0$: the corresponding  values are plotted  in Fig.~\ref{fig:RegimeS-fig2}. 

In both the stationary and oscillation regimes, the experimental variation of  $\lambda_s$ with $D/H$ is 
similar to that predicted by  Ben Richou et al.~\cite{Richou2004,Richou2005} (also plotted on the figure).
The difference between the experimental and predicted values is at most   $15\%$: 
it is likely due to inertial effects, in agreement with the variations   of the drag with the Reynolds
number reported by Hu~\cite{Hu95} and Ben Richou~\cite{Richou2005}.
Note that the influence of the space between the ends of the rod and the
 lateral sides of the cell cannot account for this difference : the corresponding  bypass flow   
would indeed instead reduce the measured value of $\lambda_s$  (see Fig.$9$ in Ref.~\onlinecite{Semin2009}).

For $D/H = 0.48$,  the transition from the stationary to the oscillation regime leads
to a small increase ($\sim 15\%$) of $\lambda_s$. This variation reflects 
the complex interplay between the rolling motion of the cylinder and its  displacement
across the gap during the oscillations. The values in the two regimes are however remarkably similar.

The second geometrical  factor $\lambda_{p}$ is determined from the
slope of the curves $V_{cx}$ vs $U$  in Fig.~\ref{fig:RegimeT-fig1}; from Eq.(\ref{eq:theory}),
 this slope must indeed be equal to $\lambda_{p}/\lambda_{s}$. Again, the experimental values
of $\lambda_p$ are higher than the theoretical ones due to inertial effects (Fig.~\ref{fig:RegimeS-fig2}).
%%%%%%%%%%%%%%%%%%%%%%%%%%%%%%%%
\section{Oscillation regime}
\label{sec:RegimeS}
%%%%%%%%%%%%%%%%%%%%%%%%%%%%%%%
%%%%%%%%%%%%%%%%%%%%%%%%%%%%
\subsection{Time variation of the transverse displacement in the gap}
%%%%%%%%%%%%%%%%%%%%%%%%%
%%%%%%%%%%%%%%%%%%%%%%%%%%%%%%%%%%%%%%%%
\begin{figure}[htbp]
\includegraphics[width=11cm]{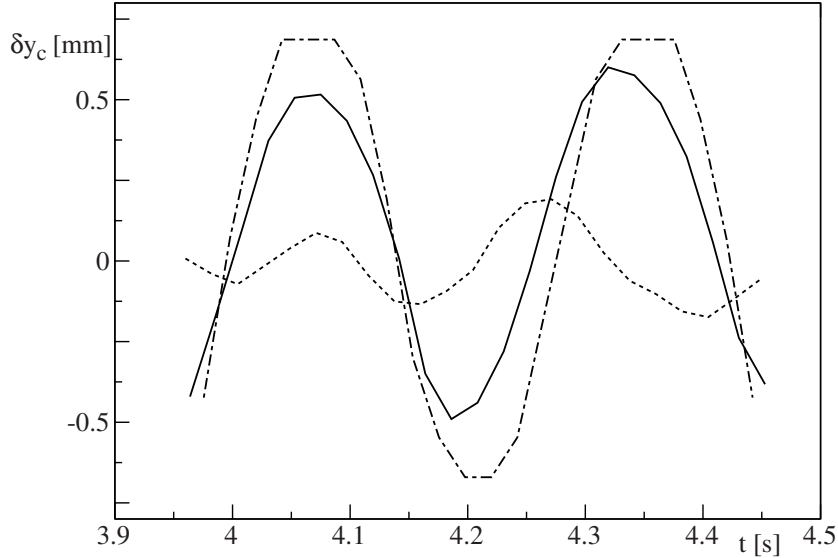} \caption{(a) variation of the transverse deviation $\delta y_c$ 
of a PMMA cylinder ($D = 1.45\ \mathrm{mm}$) in the gap as function of time for different mean  velocities $U$
of solution $N1$.
Dashed line: ($U =  - 2\ \mathrm{mm.s}^{-1}$, $Re=14$), solid line: ($U=  - 4\ \mathrm{mm.s}^{-1}$,
$Re=15$), dashed dotted line: ($U = - 12\ \mathrm{mm.s}^{-1}$, $Re=19$).}
\label{fig:figY} 
\end{figure}
%%%%%%%%%%%%%%%%%%%%%%%%%%%%%%%%%%%%%%%%
Examples of variations  with time of the transverse displacement $\delta y_{c}$ of the cylinder
from its mean position in the gap 
 are displayed in Fig.~\ref{fig:figY} for different flow velocities $U$.
The   amplitude of the oscillations close to the threshold ($U \sim - 2\ \mathrm{mm.s}^{-1}$) is  small
 ($\sim 0.3\ \mathrm{mm}$) but it  increases
rapidly with $U$ and reaches a saturation value of the order of 
$1.3 \ \mathrm{mm}$ for $U \le - 4\ \mathrm{mm.s}^{-1}$. This maximum  is close to the clearance between
the cylinders and the cell walls  ($H - D \simeq 1.55 \mathrm{mm}$): 
it corresponds then  to cylinders
coming very close to the walls during their motion as shown by the curve corresponding to $Re = 19$.  
Fig.\ref{fig:figY} also shows that the period of the oscillation is equal within $10\ \%$  for $U = - 12$ 
and $- 4 \ \mathrm{mm.s}^{-1}$ and only $25\ \%$ lower for $- 2 \ \mathrm{mm.s}^{-1}$. This variation with the
velocity is qualitatively much slower  than for vortex shedding and  tethered cylinders~(see \onlinecite{Semin2011} and 
references therein.

%%%%%%%%%%%%%%%%%%%%%%%%%%%%%%%%%%%%%%%%
\begin{figure}[htbp]
\includegraphics[width=11cm]{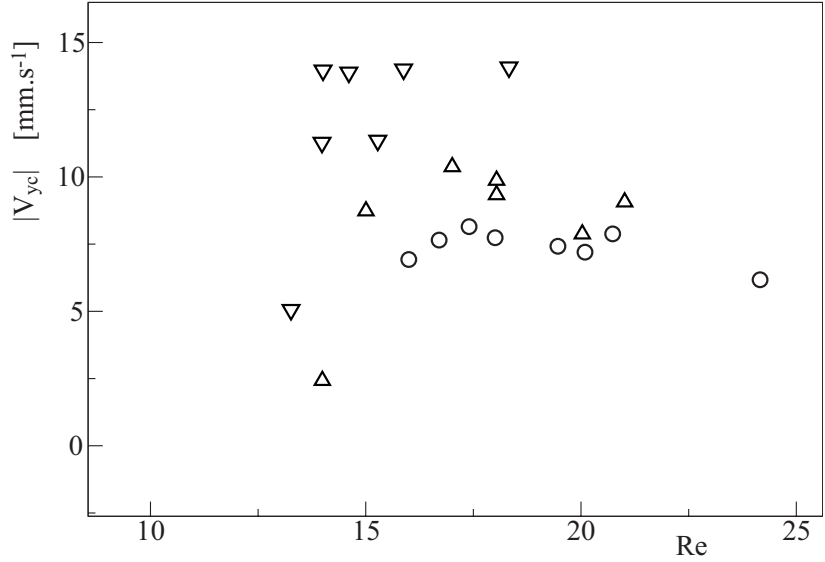} \caption{Maximum  velocity $|V_{cy}|^0$ 
of  the transverse oscillations inside the gap as a  function of the Reynolds number $Re$ defined by 
Eq.~\ref{Reynolds}. ($\triangle$), ($\circ$): 
PMMA cylinders of respective diameters $D=1.45$ and $D = 1.63\, \mathrm{mm}$ in solution $N1$. ($\triangledown$):  
carbon cylinder in solution $N2$ ($D=1.45\ \mathrm{mm}$).}
\label{fig:Vylat} 
\end{figure}
%%%%%%%%%%%%%%%%%%%%%%%%%%%%%%%%%%%
The transverse velocity $d\delta y_{c}/dt$ reaches a
maximum  $|V_{cy}^0|$  for  $\delta y_c \simeq  0$.  The  values of $|V_{cy}^0|$  measured for different cylinders 
in both solutions $N1$ and $N2$
are plotted in Fig.~\ref{fig:Vylat} as a function of the Reynolds number $Re$.
After a sharp increase over a narrow range of $Re$ values ($14 < Re < 15$),
$|V_{cy}^0|$  reaches a constant limit. This upper value decreases  as the  diameter $D$
of the cylinder increases  (($\triangle$) and ($\circ$) symbols in Fig.~\ref{fig:Vylat});  it depends also on
the cylinder density and on the fluid viscosity (($\triangle$) and ($\triangledown$) symbols). 
%%%%%%%%%%%%%%%%%%%%%%%%%%%%%%%%%%%
\subsection{Time variation of the roll angle $\alpha$.}
%%%%%%%%%%%%%%%%%%%%%%%%%%%%%%%%%%%%%%
%%%%%%%%%%%%%%%%%%%%%%%%%%%%%%%%%%%%%%%%
\begin{figure}[htbp]
\includegraphics[width=11cm]{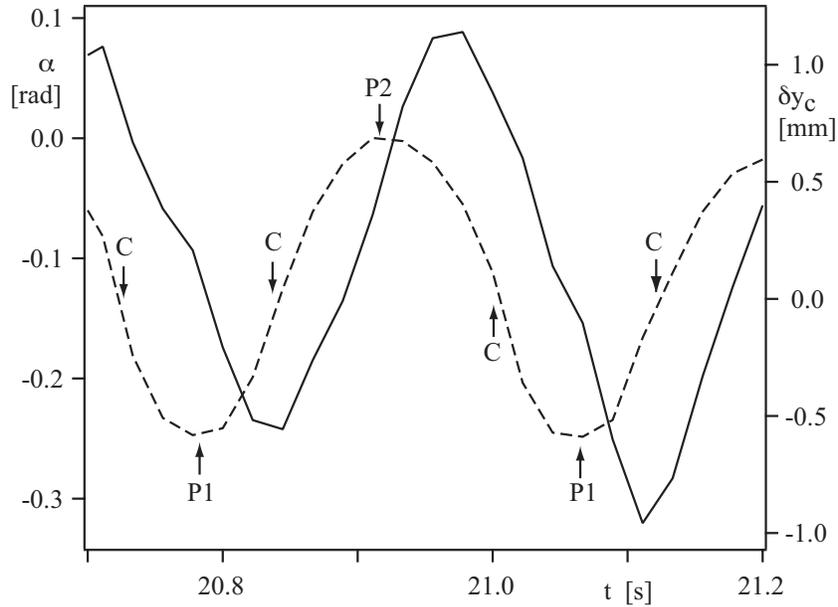} \caption{Oscillations of a PMMA cylinder 
$D=1.45\, \mathrm{mm}$ in solution $N1$ flowing at
$U = -12.7\ mm.s{-1}$ ($Re=18$). Solid line: roll angle $\alpha$; dashed line: transverse displacement $\delta y_c$.
The origin of the $\alpha$ axis is arbitrary.}
\label{fig:alpha} 
\end{figure}
%%%%%%%%%%%%%%%%%%%%%%%%%%%%%%%%%%%
The variations with time of the both the roll angle $\alpha$ and the transverse displacement
$\delta y_c$ are plotted in  Fig.~\ref{fig:alpha}. Both parameters vary periodically with the same
frequency but the shape of the variation of $\alpha$ is more triangular.
This reflects  very fast changes of the direction of the rotation which 
last for less than $0.05\ s$: they  take place shortly after the distance 
between the cylinder and the cell walls has reached its minimal value 
($\simeq 200\, \mu \mathrm{m}$).

As the cylinder moves towards one of the  walls, it rotates always in the direction opposite to the local vorticity
corresponding to the mean flow (see Fig.\ref{fig:exp}); the rotation changes direction while it moves away
so that it is again opposite to the local vorticity when it reaches the other wall.  
The corresponding absolute  tangential velocity 
$|\dot{\alpha}|D/2$ of the surface of the cylinder  at that time is  close to $9\ \mathrm{mm.s}^{-1}$ and is of the
 order of the absolute flow velocity $|U|$.  Note that the velocity of the cylinder surface facing the nearest wall has always the same sign: both the rotation direction
 and the side of the cylinder involved change indeed from one half period to the next.

%The alternate rotation of the cylinder reflects additional shear flow components (and dissipation) between 
%its surface and the walls and a large momentum exchange with the main flow: these may be expected
%to influence strongly the development of the instability.
% there is a slight shift between the data, the sign of the slope
%of the variation of $\alpha$ shifts just before
% the change of slope of $y_{c}$. 
%%%%%%%%%%%%%%%%%%%%%%%%%%%%%%%%%%%
\subsection{Variation of the vertical velocity of the particle}
%%%%%%%%%%%%%%%%%%%%%%%%%%%%%%%%%%%%
%%%%%%%%%%%%%%%%%%%%%%%%%%%%%%%%%%%%%%%%
\begin{figure}[htbp]
\includegraphics[width=11cm]{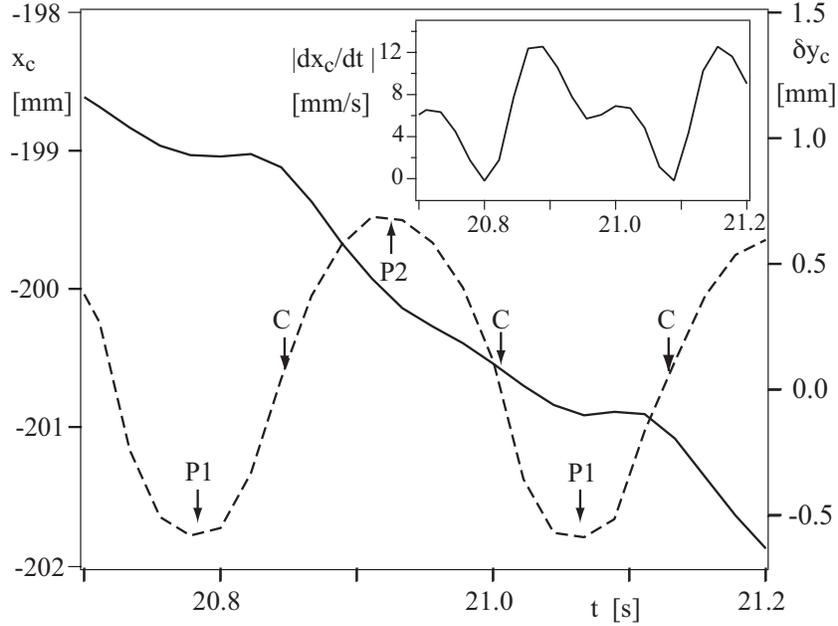} \caption{Time variation of the vertical coordinate $x_c$  (solid line)
and of the  transverse displacement $\delta y_c$ (dashed line) as a function of time for  the same experiment as in  
Fig.~\ref{fig:alpha}. Inset : variation of the absolute value vertical velocity $V_{cx} = dy_c/dt$ as a function of time.}
\label{fig:dragsed} 
\end{figure}
%%%%%%%%%%%%%%%%%%%%%%%%%%%%%%%%%%%
Figure~\ref{fig:dragsed} displays  the vertical
position of the cylinder as a function of time during three oscillations together with the
corresponding position of the particle in the gap.
In this oscillation regime, the vertical coordinate $x_c$ of the cylinder still follows a global linear trend 
with time but the velocity $V_{cx} = dx_c/dt$ displays  significant oscillations clearly visible
 in the  inset of Fig.\ref{fig:dragsed}. These  variations reflect those of the drag as the cylinder
moves across  the gap. 

The variation with time of the absolute velocity $|V_{cx}|$ displays two minima for each period of 
 the oscillation when the cylinder is near the first or the second wall (($P1$) and ($P2$) respectively).
 One of the minima is close to zero and the other much shallower, suggesting a lack of symmetry of
 the oscillation with respect to the mid-plane; this may indicate an offset of the mean transverse location
 of the cylinder from the mid plane or an asymmetry of the experimental setup.  The  absolute velocity has also two maxima
 during  each period: again, the maximum  following the lowest 
 minimum of the velocity is significantly shallower than the other. 
 %%%%%%%%%%%%%%%%%%%%%%%%%%%%%%%%%
\subsection{Variation of the oscillation frequency with the flow parameters.}
%%%%%%%%%%%%%%%%%%%%%%%%%%%%%%%%%%
%%%%%%%%%%%%%%%%%%%%%%%%%%%%%%%%%%%%%%%%
\begin{figure}[htbp]
\includegraphics[width=11cm]{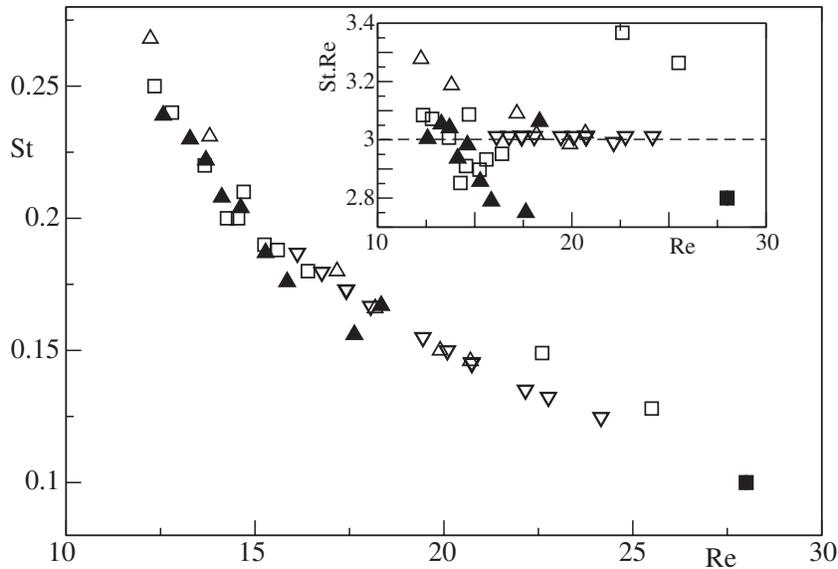}
 \caption{Variation of the Strouhal number $St = fD/U$ as a function of 
 $Re$ for different cylinders and Natrosol concentrations. Inset: variation of the product $St.Re$ as a function of $Re$. 
 Open symbols:  PMMA cylinders of diameters: ($\triangle$) $D = 1.45\, \mathrm{mm}$,  ($\triangledown$) 
 $D = 1.63\, \mathrm{mm}$,
 ($\square$) $D = 1.77\, \mathrm{mm}$ and solution $N1$; ($\blacktriangle$): carbon cylinder of diameter $D = 1.45\, \mathrm{mm}$ 
 and solution $N2$; ($\blacksquare$): PMMA cylinder of diameter $D = 1.77 \, \mathrm{mm}$ and solution $WG$.}
\label{fig:fig5} 
\end{figure}
%%%%%%%%%%%%%%%%%%%%%%%%%%%%%%%%%%%
In the oscillatory regime, it is convenient to characterize the variations of the frequency $f$ 
by  those of the  dimensionless Strouhal number:
%%%%%%%%%%%%%%%%%%%
\begin{equation}\label{strouhal}
St=\frac{f.D}{|U^{*}|}
\end{equation}
%%%%%%%%%%%%%%%%%%%%%%
Figure \ref{fig:fig5}
displays the variations of $St$ with the Reynolds number $Re$ for
all the experiments realized. Despite  the use of fluids and cylinders of 
different characteristics, all data follow a same master curve.
The value of $St$  decreases from $0.25$ to $0.1$ as $Re$ increases from
$12$ to $30$. The  inset of  figure~\ref{fig:fig5}, shows that the product $St.Re$ is
nearly constant with $Re$ and equal to $3 \pm 0.15$. The frequency $f$ of
the oscillation is then given by:
%%%%%%%%%%%%%%%%
\begin{equation}\label{freqexp} 
f=\frac{6 \nu}{D^2\displaystyle{\left(\frac{H}{D} -1\right)}} :
\end{equation} 
%%%%%%%%%%%%%%%%
with $\nu = \mu/\rho_f$.  While $f$ depends  on the fluid viscosity,  on the diameter
of the cylinder and on the ratio $H/D$, it is thefore almost independent of the  
velocity of the fluid (as  suggested above by Fig.~\ref{fig:figY}). 
%%%%%%%%%%%%%%%%%%%%%%%%%%
\section{Discussion and conclusion}
%%%%%%%%%%%%%%%%%%%%%%%%%%%%%%
The  experiments reported here  have demonstrated that cylinders free to rotate and translate in 
a vertical confined Hele Shaw cell  may display oscillations at Reynolds numbers  $10 < Re < 30$, 
well below those corresponding to vortex shedding~\cite{Williamson1996}.
These oscillations are observed over a broad range of values of the confinement
 parameter  $0.37 < D/H < 0.7$ and for different fluid viscosities $\nu$. 
 In contrast to the instability observed for tethered cylinders in a similar geometry~\cite{Semin2011}, the
 present one  induces additional  oscillations at the same frequency $f$ of both the roll angle of the cylinders
 about their axis and  their translation velocity.
 
A remarkable feature is the  weak dependence of the frequency $f$ on the  velocity $U$ while, for 
 tethered cylinders, it varies  linearly with $U$;  $f$ is in particular the same 
 for a cylinder sedimenting in a static fluid ($U = 0$)  and for an imposed  mean flow ($U \neq 0$). 
 For a given velocity  the frequency $f$ is also significantly smaller than for a tethered cylinder and
 the maximum amplitude of the oscillations of the transverse velocity is reached at a lower Reynolds 
 number (see Fig.~\ref{fig:Vylat}). One deals therefore with a very different type of instability.
 % the rotation
% of the cylinder likely plays an important part by inducing an additional shear flow component between
% its surface and the flow cell walls.

 Quantitatively, $f$  is related to $U$, $D$ and $H/D$ by Eq.~(\ref{freqexp}). This formula implies that 
$f$ is    proportional to the inverse of the characteristic viscous diffusion time  $\tau_{d} = D^2/\nu$ 
over the diameter $D$ of the cylinder;  the proportionality coefficient $(H/D - 1)/6$ accounts 
for the blockage of the flow by the cylinder.  In terms of dimensionless 
variables, the Strouhal number $St$ is not constant but increases from 
$0.1$ to $0.25$ with $Re$ (Fig.~\ref{fig:fig5}); for tethered cylinders, instead, $St$ varies  slowly
and is significantly higher ($0.65 \le St \le 0.75$). Here, the relevant combination is the product
$St Re$ of the Strouhal and Reynolds numbers which is proportional to $1/\tau_d$ and constant with $Re$. 
 
These results have direct implications on the transverse velocity in the gap at large  amplitudes: assuming
a constant absolute value of $|V_{cy}|$ during the oscillations, it can be estimated by  $|V_{cy}| \simeq 2 f (H - D)$. Using Eq.~(\ref{freqexp}), this leads to:
%%%%%%%%%%%%%%%%
\begin{equation}\label{vitexp} 
|V_{cy}|  \simeq  \displaystyle{\frac{6 \nu}{D}}.
\end{equation} 
%%%%%%%%%%%%%%%%
The velocity $|V_{cy}|$ is then also independent of the mean flow velocity (and, therefore, of the Reynolds 
number for a given viscosity $\nu$): this is indeed observed in Fig.\ref{fig:Vylat} for $Re \ge 15$. Eq.~\ref{freqexp}
also predicts that $|V_{cy}|$ increases with $\nu$ and decreases as the diameter $D$ increases,
also in agreement with the data plotted in Fig.~\ref{fig:Vylat}.

%\cor{Similarly, assuming a constant absolute value of the angular velocity during the oscillations (like in Fig.~\ref{fig:alpha}) 
%leads to a tangential velocity component at the surface of the cylinder 
%$|V_{c\theta}| \simeq  D f \Delta \alpha$ ($\Delta \alpha$ = 
%amplitude of the angular oscillations). Using  
%Eq.~\ref{freqexp}  leads to:
%%%%%%%%%%%%%%%%
%\begin{equation}\label{vithetaexp} 
%|V_{c\theta}| \simeq = \frac{6 \nu \, \Delta \alpha}{H - D}.
%\end{equation} 
%%%%%%%%%%%%%%%%
%Since the experiments indicate that $\Delta \alpha$ varies little with the viscosity $\nu$,
%$|V_{c\theta}|$ must then also be proportional to   $\nu$.}
The  variations with time of the transverse  displacement in the gap and  of the rolling angle 
are antisymmetrical with respect to the midplane:  the absolute transverse
and rotational velocities are 
 the same when  the cylinder moves towards a wall or away from it.
This symmetry is not as well satisfied by the vertical velocity component (and therefore by
the vertical drag force on the cylinder).

The above results suggest that the growth of the oscillations of the cylinder is 
determined by the relative values of their period $1/f$ and of $\tau_d$. At low $f$ values ($\ll 1/\tau_d$),
momentum diffusion is fast enough so that the flow field around  the cylinder  reaches 
a quasistatic profile identical to that of a fixed cylinder at the same location. 
 At high frequencies ($f \gg 1/\tau_d$), there is a phase shift 
between the motion of the cylinder and the corresponding variations of the flow and pressure fields
(particularly in quasiparallel flow regions in the gaps between the cylinder and the walls).
At Reynolds numbers large enough (here $\ge \, 14$) so that non linear terms appear, the resulting force 
on the cylinder may then change  sign due to the phase shift and amplify the oscillations instead of damping them.

%As noted above, an important feature is the very weak influence on the frequency of the oscillations of a \coc{mean} flow superimposed
%over the sedimentation of the cylinder.
%Actually the influence of these two components may be accounted for globally
%by using the single combined velocity $\bf{U}^*$. 
%At low Reynolds numbers, instead, different behaviours have been reported 
%for  cylinders  sedimenting  \cite{Dvinsky87a,Hu95} or  transported by
%a \coc{mean} flow \cite{Dvinsky87b,Eklund1994}. 
 
At the largest values of the confinement parameter $D/H$,  more complex
dynamical phenomena were observed. In addition to the instability reported above, 
the cylinder displayed flutter with  lower frequency oscillations
 of its angle with respect to the horizontal and of its lateral position. 
Numerous studies have analyzed similar motions involving the coupling between the vertical
motion of   objects  and lateral oscillations: these are encountered in such problems as the 
fluttering motion of falling leaves or paper sheets~\cite{Tanabe1994,Belmonte1998} or that of 
bubbles rising in a liquid~\cite{Magnaudet2000}. It will be necessary 
in future work to quantify the effect of the confinement on these observations and consider the
low Reynolds limit. 

Finally, we considered only cylinders of length $L_c$ close to  the 
width $W$ of the cell. With shorter cylinders for which $L_c < W$, the bypass
of the flow between the tips of the cylinders and the lateral walls
will reduce the influence of the flow blockage and thus influence
the motion. This, too, represents an important parameter of the problem.

%%%%%%%%%%%%%%%%%%%%%%%%%%%%%%%%%%%%%%%%%%
\begin{acknowledgements} %%%%%%%%%%%%%%%%%%%%%%%%%%%%%%%%%%%%%%%%%%
We thank R. Pidoux, L. Auffay and A. Aubertin for realizing and developing
the experimental set up and B. Semin for his careful reading of the manuscript and his useful comments. 
We acknowledge the RTRA Triangle de la Physique
and the LIA PMF-FMF (Franco-Argentinian International Associated Laboratory
in the Physics and Mechanics of Fluids). The work of one of us (VD) was supported by
a Bernardo Houssay grant allocated by the Argentinian and French
ministries of research. 
%%%%%%%%%%%%%%%%%%%%%%%%%
\end{acknowledgements} 
%%%%%%%%%%%%%%%%%%%%%%%%%%%%%%%%%

% Create the reference section using BibTeX:
%\bibliography{vero26.bbl}
%%%%%%%%%%%%%%%%%%%%%%%%%%%%%%%%%%%%%%%%%%%%%%%%

%%%%%%%%%%%%%%%%%%%%%%%%%%%%%%%%%%%%%%%%%%%%%%%%
%\bibitem{}
% and use \bibitem to create references.
%\bibitem{RefJ}
% Format for Journal Reference
%Author, Journal \textbf{Volume}, (year) page numbers.
% Format for books
%\bibitem{RefB}
%Author, \textit{Book title} (Publisher, place year) page
% etc

\end{document}